\RequirePackage{lineno}
\documentclass[floatfix,superscriptaddress,a4paper,
               showpacs,showkeys,nofootinbib,preprint]{revtex4}
\usepackage{epsfig}
\usepackage{latexsym}
\usepackage{xspace}

\usepackage{inputenc}
\usepackage{indentfirst}
\usepackage{enumerate}
\usepackage{color}
\usepackage{epstopdf}
\usepackage{colortbl}
\usepackage{blindtext}
\usepackage{lineno}
\usepackage{amsmath}
\usepackage{amssymb}
\usepackage[english]{babel}
\usepackage{url}
\usepackage{hyperref}

\graphicspath{{plots/}}

\begin{document}

%%%%%%% units
\newcommand{\kg}{\ensuremath{\mbox{kg}}\xspace}
\newcommand{\eV}{\ensuremath{\mbox{e\kern-0.1em V}}\xspace}
\newcommand{\GeV}{\ensuremath{\mbox{Ge\kern-0.1em V}}\xspace}
\newcommand{\MeV}{\ensuremath{\mbox{Me\kern-0.1em V}}\xspace}
\newcommand{\GeVc}{\ensuremath{\mbox{Ge\kern-0.1em V}\!/\!c}\xspace}
\newcommand{\GeVcc}{\ensuremath{\mbox{Ge\kern-0.1em V}\!/\!c^2}\xspace}
\newcommand{\AGeV}{\ensuremath{A\,\mbox{Ge\kern-0.1em V}}\xspace}
\newcommand{\AGeVc}{\ensuremath{A\,\mbox{Ge\kern-0.1em V}\!/\!c}\xspace}
\newcommand{\MeVc}{\ensuremath{\mbox{Me\kern-0.1em V}/c}\xspace}
\newcommand{\T}{\ensuremath{\mbox{T}}\xspace}
\newcommand{\cmsq}{\ensuremath{\mbox{cm}^2}\xspace}
\newcommand{\msq}{\ensuremath{\mbox{m}^2}\xspace}
\newcommand{\cm}{\ensuremath{\mbox{cm}}\xspace}
\newcommand{\mm}{\ensuremath{\mbox{mm}}\xspace}
\newcommand{\micron}{\ensuremath{\mu\mbox{m}}\xspace}
\newcommand{\mrad}{\ensuremath{\mbox{mrad}}\xspace}
\newcommand{\ns}{\ensuremath{\mbox{ns}}\xspace}
\newcommand{\m}{\ensuremath{\mbox{m}}\xspace}
\newcommand{\s}{\ensuremath{\mbox{s}}\xspace}
\newcommand{\ms}{\ensuremath{\mbox{ms}}\xspace}
\newcommand{\ps}{\ensuremath{\mbox{ps}}\xspace}
\newcommand{\dd}{\ensuremath{{\textrm d}}\xspace}
\newcommand{\dedx}{\ensuremath{\dd E\!/\!\dd x}\xspace}
\newcommand{\tof}{\ensuremath{\textup{\emph{tof}}}\xspace}
\newcommand{\pt}{\ensuremath{p_{\textrm T}}\xspace}
\newcommand{\PT}{\ensuremath{P_\textup{T}}\xspace}
\newcommand{\mt}{\ensuremath{m_{\textrm T}}\xspace}

%particles
\newcommand{\pbar}{\ensuremath{\overline{\textup{p}}}\xspace}
\newcommand{\p}{\ensuremath{\textup{p}}\xspace}
\newcommand{\nbar}{\ensuremath{\overline{\textup{n}}}}
\newcommand{\dbar}{\ensuremath{\overline{\textup{d}}}}
\newcommand{\pim}{\ensuremath{\pi^-}\xspace}
\newcommand{\pip}{\ensuremath{\pi^+}\xspace}
\newcommand{\km}{\ensuremath{\textup{K}^-}\xspace}
\newcommand{\kp}{\ensuremath{\textup{K}^+}\xspace}
\newcommand{\hm}{\ensuremath{\textup{h}^-}\xspace}

% inverse hyperbolic functions
%\DeclareMathOperator{\acosh}{acosh}
%\DeclareMathOperator{\asinh}{asinh}
%\DeclareMathOperator{\atanh}{atanh}

%%%%%%%%%%%%% some software programs and generators
%----- NA61 software
\def\Offline{\mbox{$\overline{\text%
{Off}}$\hspace{.05em}\raisebox{.4ex}{\underline{line}}}\xspace}
\def\SHOE{\mbox{SHO\hspace{-1.34ex}\raisebox{0.2ex}{\color{green}\textasteriskcentered}\hspace{0.25ex}E}\xspace}
\def\DSHACK{\mbox{DS\hspace{0.15ex}$\hbar$ACK}\xspace}
\DeclareRobustCommand{\SHINE}{\mbox{\textsc{S\hspace{.05em}\raisebox{.4ex}{\underline{hine}}}}\xspace} %DeclareRobustCommand allows this to work in caption
%\def\SHINE{\textsc{Shine}\xspace}
%----- event generators
\def\Glissando{\textsc{Glissando}\xspace}
\newcommand{\FlukaLong}{{\scshape Fluka2008}\xspace}
\newcommand{\FlukaEleven}{{\scshape Fluka2011}\xspace}
\newcommand{\Fluka}{{\scshape Fluka}\xspace}
\newcommand{\UrqmdLong}{{\scshape U}r{\scshape qmd1.3.1}\xspace}
\newcommand{\Urqmd}{{\scshape U}r{\scshape qmd}\xspace}
\newcommand{\GheishaLong}{{\scshape Gheisha2002}\xspace}
\newcommand{\GheishaOld}{{\scshape Gheisha600}\xspace}
\newcommand{\Gheisha}{{\scshape Gheisha}\xspace}
\newcommand{\Corsika}{{\scshape Corsika}\xspace}
\newcommand{\Venus}{{\scshape Venus}\xspace}
\newcommand{\VenusLong}{{\scshape Venus4.12}\xspace}
\newcommand{\GiBUU}{{\scshape GiBUU}\xspace}
\newcommand{\GiBUULong}{{\scshape GiBUU1.6}\xspace}
\newcommand{\FlukaNewLong}{{\scshape Fluka2011.2\_17}\xspace}
\newcommand{\Root}{{\scshape Root}\xspace}
\newcommand{\Geant}{{\scshape Geant}\xspace}
\newcommand{\GeantThree}{{\scshape Geant3}\xspace}
\newcommand{\GeantFour}{{\scshape Geant4}\xspace}
\newcommand{\QGSJet}{{\scshape QGSJet}\xspace}
\newcommand{\DPMJet}{{\scshape DPMJet}\xspace}
\newcommand{\Epos}{{\scshape Epos}\xspace}
\newcommand{\EposLong}{{\scshape Epos1.99}\xspace}
\newcommand{\QGSJetLong}{{\scshape QGSJetII-04}\xspace}
\newcommand{\DPMJetLong}{{\scshape DPMJet3.06}\xspace}
\newcommand{\SibyllLong}{{\scshape Sibyll2.1}\xspace}
\newcommand{\EposLHCLong}{{\scshape EposLHC}\xspace}
\newcommand{\Hsd}{{\scshape Hsd}\xspace}
\newcommand{\Ampt}{{\scshape Ampt}\xspace}

\newcommand{\NASixtyOne}{NA61\slash SHINE\xspace}

%%%%%%%%%%%%%%%%%%%%%%%% misc
\def\red#1{{\color{red}#1}}
\def\blue#1{{\color{blue}#1}}
\def\avg#1{\langle{#1}\rangle}
\def\sci#1#2{#1\!\times\!10^{#2}}

\title{Diagram of High-Energy Nuclear Collisions\footnote{This paper is based on work performed before February 24th 2022.}}

\author{Evgeny Andronov}
\affiliation{St. Petersburg State University, St. Petersburg, Russia}

\author{Magdalena Kuich}
\affiliation{University of Warsaw, Warsaw, Poland}

\author{Marek Gazdzicki}
\affiliation{Geothe-University Frankfurt am Main, Germany}
\affiliation{Jan Kochanowski University, Kielce, Poland}

\begin{abstract}

Many new particles, mostly hadrons, are produced in high-energy collisions between atomic nuclei.
The most popular models describing the hadron-production process are
based on the creation, evolution and decay of  resonances, strings or quark--gluon plasma.
The validity of these models is under vivid discussion, and it seems that a common framework for this discussion is missing.
Here, for the first time, we explicitly introduce the diagram of high-energy nuclear collisions, where domains of the dominance of different hadron-production processes in the space of laboratory-controlled parameters, the collision energy and nuclear-mass number of colliding nuclei are indicated.
We argue that
the recent experimental results suggest the location of boundaries between 
the domains, allowing for the first time to sketch an example diagram.
Finally, we discuss the immediate implications for experimental measurements and model development following the proposed sketch of the diagram.

\end{abstract}

\pacs{25.75.q, 25.75.Nq, 24.60.Ky}

\keywords{high-energy collisions, strongly interacting matter, quark-gluon plasma, strings, resonances}
\maketitle

%\linenumbers

\section{Introduction}
\label{sec:introduction}

One of the crucial issues of contemporary physics is understanding  
strong interactions - the interactions defining properties of atomic nuclei and collisions between them. Nuclear collisions at high energies lead to the production of many new particles, predominately strongly interacting hadrons. 
With the advent of the quark model of hadrons and
the development of the commonly accepted theory of strong interactions, 
quantum chromodynamics (QCD) naturally led to expectations that matter at very high densities may exist in a state of quasi-free quarks and gluons, the~quark--gluon plasma (QGP)~\cite{Shuryak:1980tp}. There are numerous indications that QGP is created in heavy-ion collisions at high energies; for~review, see Refs.~\cite{Heinz:2000bk, STAR:2005gfr, Gazdzicki:2010iv, Gazdzicki:2020jte}. 

The theoretical description of high-energy nuclear collision is not an easy task.
This may be attributed to the difficulty of obtaining unique and quantitative predictions from QCD. 
In particular, even the formation of QGP in heavy-ion collisions is beyond the predictability of QCD.
Consequently, the~bulk properties of high-energy nuclear collisions are described by phenomenological models. 
Over time, three classes of them gained~in popularity:
\begin{enumerate}
\item [(i)] One postulates that the final hadronic state emerges from the quark--gluon plasma's creation, evolution, and~ hadronisation~\cite{Florkowski:2010zz}.
A key input - the QGP equation of state - can be estimated using lattice-QCD calculations~\cite{Karsch:2001cy}.
This process will be labelled as \emph{QGP};

\item [(ii)]
One assumes hadrons originate from the formation, evolution and fragmentation of strings - the gluon fields between a pair of colour charges forming a narrow flux tube~\cite{Werner:1993uh}. 
Strings are typically oriented along the collision axis, and~they have a continuous masses spectrum. Symmetries and experimental results are used to determine model parameters.
This process will be labelled as \emph{strings};
\item [(iii)]
One describes the production of final state hadrons by creation, evolution and decay of hadronic resonances~\cite{Bleicher:1999xi} - excited states of stable hadrons. 
Resonances do not have a preferred elongation direction and have a discrete mass spectrum. 
Experimental results are used to determine model parameters. 
This hadron-production process will be
labelled as \emph{resonances}.
\end{enumerate}

Do the processes reflect reality? 
If yes, what are the domains of their applicability? 
Answering these questions is directly related to
understanding intriguing changes of hadron-production properties observed experimentally by varying collision energy and the mass number of colliding nuclei. 
This task goes hand in hand with selecting measurable quantities sensitive 
to a transition between the processes.
In this paper, we focus the discussion on the ratio of positively charged kaons and pions
measured at mid-rapidity, the~$K^+/\pi^+$ ratio.
This measure can be interpreted as a good approximation of the strange to non-strange quarks ratio.
Due to mass and number differences between strange and non-strange particles (quarks and gluons or hadrons), the~ratio is expected to be sensitive to the hadron-production process~\cite{Rafelski:1982pu,Gazdzicki:1998vd} - it is expected to be sensitive to a changeover between
different processes. With~the above and the availability of the rich experimental data, 
the choice of the $K^+/\pi^+$ ratio as the subject of this paper was most~suitable.

For a quantitative comparison of the experimental results with model predictions, we selected PHSD~\cite{Cassing:2008sv,Cassing:2009vt} and SMASH~\cite{Weil:2016zrk,Mohs:2019iee} models. This is motivated by their important features. Both models give predictions in the full range
collision energy and masses of the colliding nuclei covered by the experimental data. The~SMASH model includes resonances and strings, whereas the PHSD model also includes~QGP.

We review the experimental results and suggest the first answers to the 
questions asked in Section~\ref{sec:data}. 
Section~\ref{sec:diagram} introduces
the diagram of high-energy nuclear collisions, and~we summarise our findings in a diagram sketch.
Finally, we discuss the implications following the sketch for experimental measurements and developing~models.

\section{Guiding Ideas and Experimental~Results}
\label{sec:data}

%\vspace{0.2cm}
\textbf{Heavy-ion collisions.}

\begin{figure}[h]
	%\centering
	\includegraphics[width=0.49\textwidth]{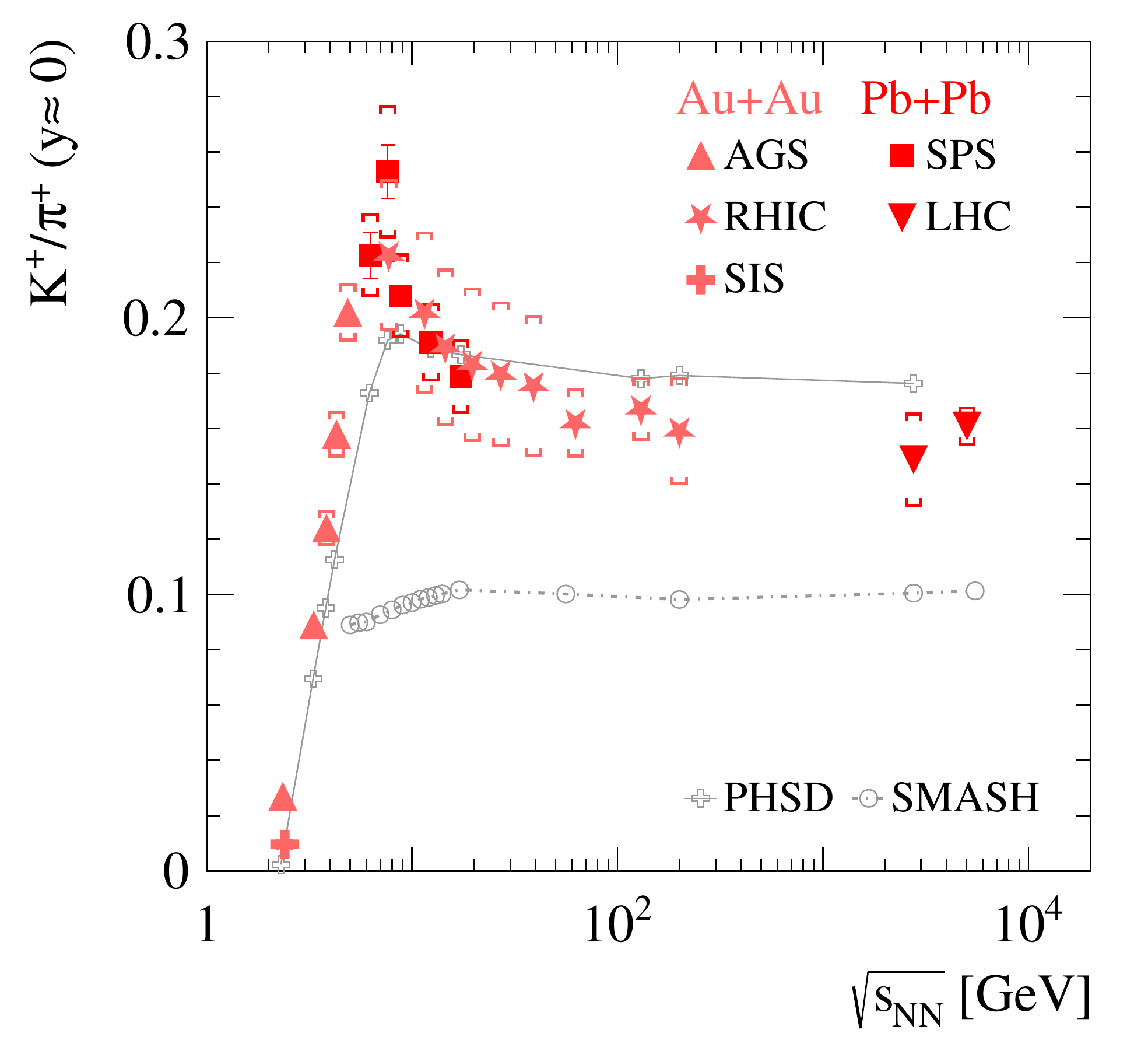}
	\includegraphics[width=0.49\textwidth]{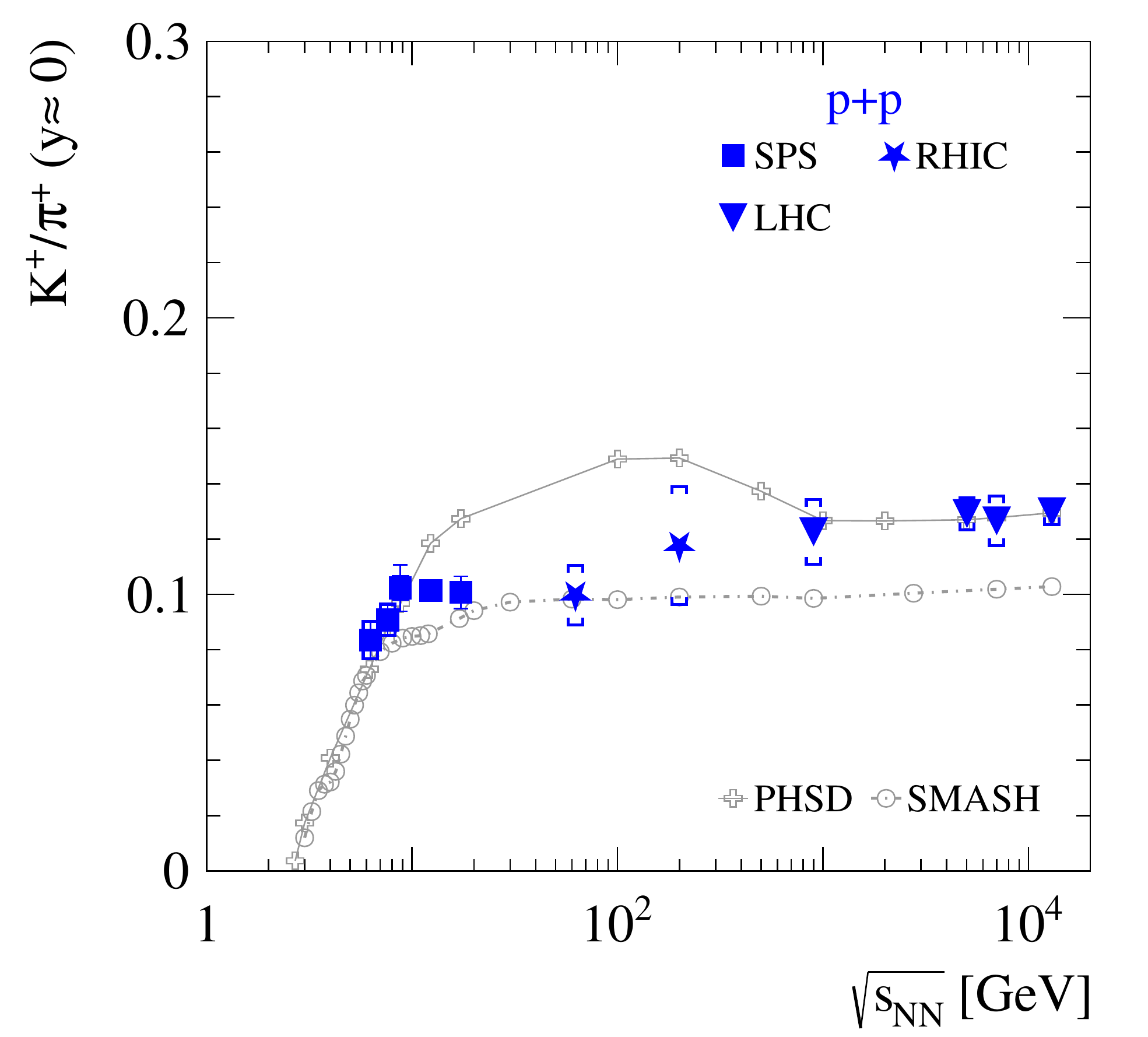}
	\caption{Collision energy dependence of the $K^+/\pi^+$ multiplicity ratio at mid-rapidity in central heavy-ion collisions (Pb+Pb~\cite{NA49:2007stj,NA49:2002pzu,PhysRevC.88.044910,ALICE:2019hno} and Au+Au~\cite{E866:1999ktz,E0895:2003oas,STAR:2017sal,STAR:2019vcp,STAR:2008med,HADES:2018adm,HADES:2020ver})~(\textbf{left}) and in inelastic \emph{p+p} interactions~\mbox{\cite{NA61SHINE:2017fne,PHENIX:2011rvu,ALICE:2011gmo,ALICE:2015ial,ALICE:2019hno,ALICE:2020jsh}}~(\textbf{right}). Open cross points present the PHSD~\cite{Cassing:2008sv,Cassing:2009vt} predictions, while open circles - the SMASH~\cite{Weil:2016zrk,Mohs:2019iee} predictions. Lines connecting the points are plotted to guide the eye.
	}
	\label{fig:horn}
\end{figure}

The richest experimental results on the collision energy dependence of hadron-production properties concern collisions between two heavy atomic nuclei, Pb+Pb and Au+Au collisions.
Over the last 40 years, they were recorded in the hunt for QGP and the energy threshold of its creation - the \emph{onset of deconfinement}. Many fixed-target and collider 
experiments in the US (Lawrence Berkeley Laboratory, LBL, and~Brookhaven National Laboratory, BNL) and European (European Organization of Nuclear Research, CERN and Helmholtz Centre for Heavy Ion Research, GSI) laboratories have been conducting the measurements.
The results are consistent with the onset of deconfinement being located at 
($\sqrt{s_{NN}}\approx 8$~GeV) and the QGP being created at the early stage of heavy-ion
collisions at higher collision energies (for review, see Refs.~\cite{Heinz:2000bk, STAR:2005gfr, Gazdzicki:2010iv, Gazdzicki:2020jte}).
The most popular plot illustrating this assessment is presented in Figure~\ref{fig:horn}~(left).
It shows the collision energy dependence of the $K^+/\pi^+$ ratio in central heavy-ion collisions.
The ratio shows the so-called \emph{horn} structure. Following a fast rise, the~ratio passes through a maximum in the CERN SPS energy range, at~approximately 8~GeV, then decreases and settles to a plateau which continues up to the CERN LHC energies.
Kaons are the lightest strange hadrons, and~due to approximate isospin symmetry, the 
$K^+$ yield counts about half of the strange quarks produced in the collisions and contained in the reaction products~\cite{Gazdzicki:1998vd}. 
Thus, Figure~\ref{fig:horn}~(left) demonstrates
that the fraction of strangeness carrying particles in the produced matter passes through
a sharp maximum at the SPS energy range in central heavy-ion collisions;
for a detailed explanation; see Ref.~\cite{Gazdzicki:2010iv}.

The standard modelling of heavy-ion collisions~\cite{Florkowski:2010zz} includes the formation of high-density matter (be it QGP or hadronic matter) at the early stage of a collision, its expansion and the decoupling of hadrons that freely stream to particle detectors. A~statistical description of the early stage~\cite{Gazdzicki:1998vd} led to predictions of the collision energy dependence of bulk hadron production properties.
In particular, the~horn structure was predicted as the signal of the 
onset of deconfinement. In~the model, it reflects the decrease in the ratio of strange to non-strange degrees of freedom when deconfinement sets in.
Experimental data are compared with calculations of the PHSD model~\cite{Cassing:2008sv,Cassing:2009vt} that incorporates the QGP creation at sufficiently high densities and chiral-symmetry restoration in the dense hadronic matter.
The model catches the basic properties of the data; 
see Figure~\ref{fig:horn}~(left).  
This further supports interpreting the horn maximum  
at $\sqrt{s_{NN}}\approx 8$~GeV as the beginning of the QGP creation.
Moreover, the~SMASH model~\cite{Weil:2016zrk,Mohs:2019iee}, which does not include the QGP creation qualitatively, fails to reproduce the results;
see Figure~\ref{fig:horn}~(left). 
One should, however, note that there are significant uncertainties in modelling both production processes; see below for an~example.

%\vspace{0.2cm}
\textbf{Proton--proton interactions.}

Measurements of proton--proton interactions started long before the first experiments studying heavy-ion collisions.
The primary goal of the study of \emph{p+p} interactions was understanding strong interactions.
With increasing collision energy, more and heavier hadrons have been produced.
To understand the early results, the~string model was invented~\cite{Veneziano:1968yb}.
The experimental results also contributed to the formulation of QCD - the nowadays commonly accepted theory of strong interactions. 
While there is no first-principles derivation of strings from QCD, some properties of a string can be derived from~QCD.

Paradoxically, QCD demotivated studies of bulk properties of \emph{p+p} interactions.
This is because of difficulties in obtaining unique and quantitative predictions 
from QCD.
Consequently, the~world data on \emph{p+p} interactions are not as rich as the
corresponding results on heavy-ion collisions.
The compiled data on the $K^+/\pi^+$ ratio at mid-rapidity in
inelastic \emph{p+p} interactions is shown in
Figure~\ref{fig:horn}~(right).
Precise measurements are available at the CERN SPS and LHC energies.
The \emph{p+p} results still allow for important~observations:
\begin{enumerate}
\item [(i)]
At the SPS energies, the~ratio in \emph{p+p} interactions is about a factor of two lower than in heavy-ion collisions;
\item [(ii)]
At $\sqrt{s_{NN}}\approx 8$~GeV, a~\emph{break} in the collision energy dependence of the ratio is observed in \emph{p+p} interactions instead of the horn seen in heavy-ion collisions. For~a more detailed analysis of the \emph{p+p} break, see Ref.~\cite{NA61SHINE:2019xkb};
\item [(iii)]
At the LHC energies, the~\emph{p+p} ratio is about 20\% lower than the heavy-ion one.
\end{enumerate}

The most popular modelling of proton--proton interactions at high energies
includes strings' formation, evolution, and~fragmentation. The~widely used approaches are the Lund~\cite{Andersson:1983ia}, EPOS~\cite{Werner:1993uh} and Dual Parton~\cite{Kaidalov:1982xe,Capella:1992yb} models.  
At low collision energies, the~validity of the string approach breaks, and~one replaces it with the creation of resonances and their decay; 
for a detailed explanation, see Ref.~\cite{Mohs:2019iee}. These two processes are implemented in the
PHSD~\cite{Cassing:2008sv,Cassing:2009vt} and SMASH~\cite{Weil:2016zrk,Mohs:2019iee} models. 

Their predictions for the collision energy dependence of the $K^+/\pi^+$ ratio in \emph{p+p} interactions
are shown in Figure~\ref{fig:horn}~(right).
Significant differences between them shed light on the uncertainty of the
predictions. 
Taking into account this uncertainty, one concludes that
the models reproduce the bulk properties of the~data.

The effect of the changeover from resonances to strings 
(\emph{onset of strings})
was studied in detail within the UrQMD model~\cite{Bass:1998ca, Vovchenko:2014vda}.
Within SMASH~\cite{Weil:2016zrk,Mohs:2019iee}, the~changeover causes a wiggle in the collision energy dependence of the $K^+/\pi^+$ ratio, which can be seen in Figure~\ref{fig:horn}~(right) by enlarging the plot.
In PHSD~\cite{Cassing:2008sv,Cassing:2009vt}, a~sharp transition is located at $\sqrt{s_{NN}} \approx 2.6$~GeV - close to the threshold for kaon production---and thus its effect on the ratio is hard to~observe.

The open question discussed in Ref.~\cite{NA61SHINE:2019xkb} is whether the break (ii) in the collision energy dependence of the experimental ratio at $\sqrt{s_{NN}} \approx 8$~GeV is due to the onset of strings or is related to the onset of~deconfinement.

One notes the following regarding the similarity of the ratio in \emph{p+p} and Pb+Pb collisions at LHC (iii).
It was reported that relative strange hadron yields 
in \emph{p+p} interactions at LHC smoothly increase 
with increasing charged-particle multiplicity and for high multiplicity interactions are close to those in Pb+Pb collisions~\cite{ALICE:2017jyt}.
Moreover, recent LHC data on the azimuthal angle distribution of charged particles in high multiplicity \emph{p+p} interactions~\cite{CMS:2012qk,Khachatryan:2016txc,Nagle:2018nvi} show anisotropies up to the recently observed
only in heavy-ion collisions and attributed to the hydrodynamical expansion of matter~\cite{Bozek:2016acp}. 
This suggests that QGP 
may also be produced in \emph{p+p} interactions at the LHC energies, at~least in collisions with sufficiently high hadron~multiplicity.\vspace{-3pt}

%\vspace{0.2cm}
\textbf{Collisions of intermediate-mass nuclei.}
%mkuich: please keep this format

The collision-energy dependence of hadron-production properties in
collisions of intermediate-mass nuclei is the least established one.
The only systematic measurements have been performed at the CERN SPS
by NA61/SHINE~\cite{Abgrall:2014xwa}.

\begin{figure}[h!]
	%\centering
	\includegraphics[width=0.69\textwidth]{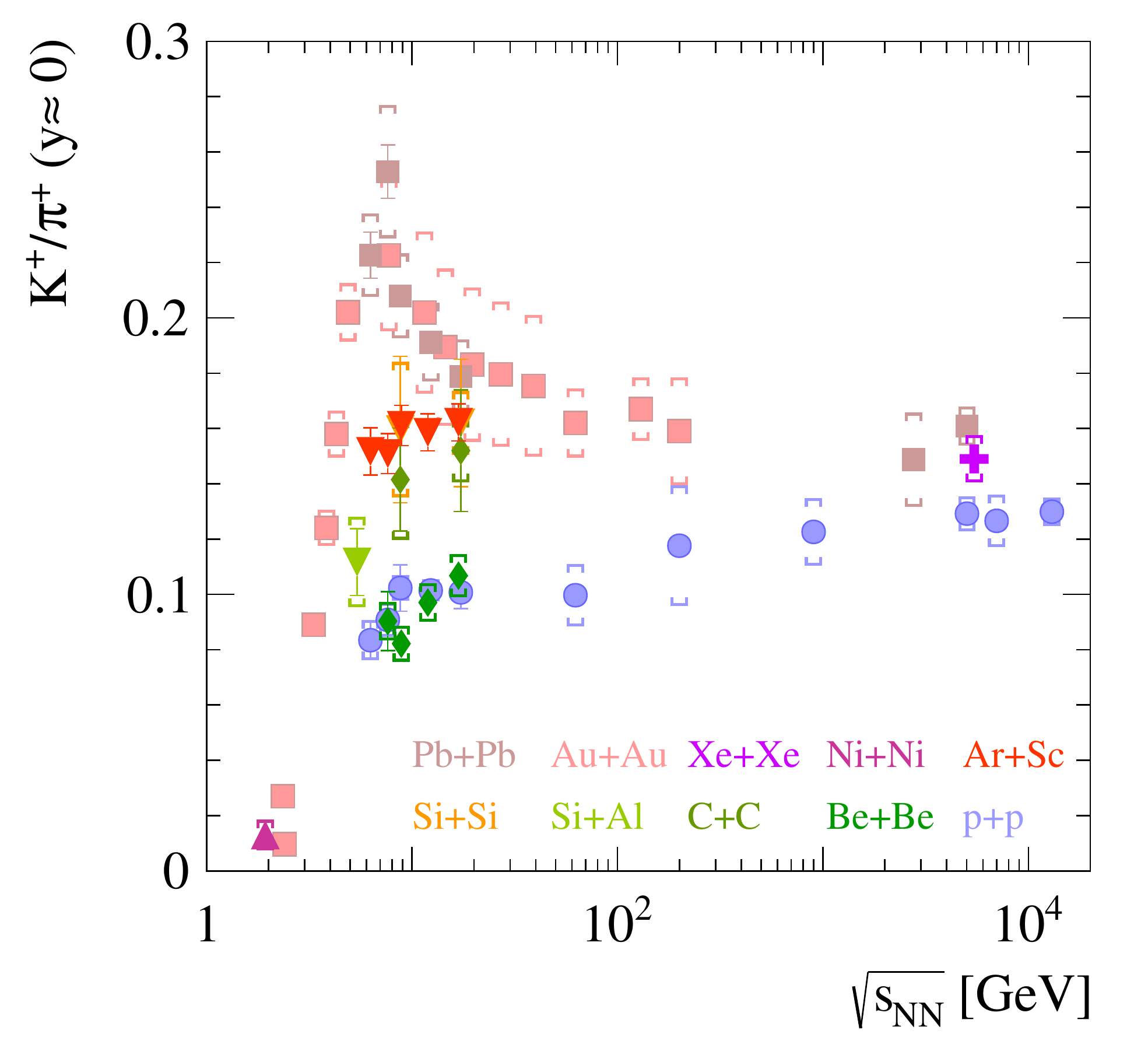}
	\caption{Status of experimental results on energy dependence of the $K^+/\pi^+$ ratio at mid-rapidity in high-energy nuclear collisions. In~addition to the previously shown results on central heavy-ion collisions~\cite{NA49:2007stj,NA49:2002pzu,PhysRevC.88.044910,ALICE:2019hno,E866:1999ktz,E0895:2003oas,STAR:2017sal,STAR:2019vcp,STAR:2008med,HADES:2018adm,HADES:2020ver} and inelastic \emph{p+p}~\cite{NA61SHINE:2017fne,PHENIX:2011rvu,ALICE:2011gmo,ALICE:2015ial,ALICE:2019hno,ALICE:2020jsh} interactions, results on central Be+Be~\cite{NA61SHINE:2020czq}, C+C~\mbox{\cite{NA49:2004jzr,NA49:2012rsi}, Si+Al~\cite{E-802:1994hea}}, Si+Si~\cite{NA49:2004jzr,NA49:2012rsi}, Ar+Sc~(preliminary)~\cite{Lewicki:2020mqr,NA61SHINE:2021nye}, Ni+Ni~\cite{FOPI:1997rgs,FOPI:1997tif} and Xe+Xe~\cite{ALICE:2021lsv} collisions are shown. Points presenting Pb+Pb/Au+Au and \emph{p+p} are plotted in pale colours to emphasize intermediate-mass nuclei results.
	}
	\label{fig:intermidiate}
\end{figure}

They were motivated by a search for the critical point of strongly interacting matter and a need to establish the nuclear mass dependence
of the horn structure~\cite{Gazdzicki:995681}. The~data on collisions of intermediate-mass nuclei are summarised in Figure~\ref{fig:intermidiate}.
The results of $K^+/\pi^+$ ratio for central Pb+Pb/Au+Au and inelastic \emph{p+p} are also plotted for comparison in a light colour.
The main observations~are:
\begin{enumerate}
\item [(i)]
The ratio in Be+Be collisions is similar to the one in \emph{p+p} interactions in the whole SPS energy range;
\item [(ii)]
There is no horn structure in Ar+Sc collisions;
\item [(iii)]
The ratio in Ar+Sc collisions at the top SPS energy is similar to
the one in Pb+Pb collisions.
\end{enumerate}

\begin{figure}[h]
	%\centering
	\includegraphics[width=0.49\textwidth]{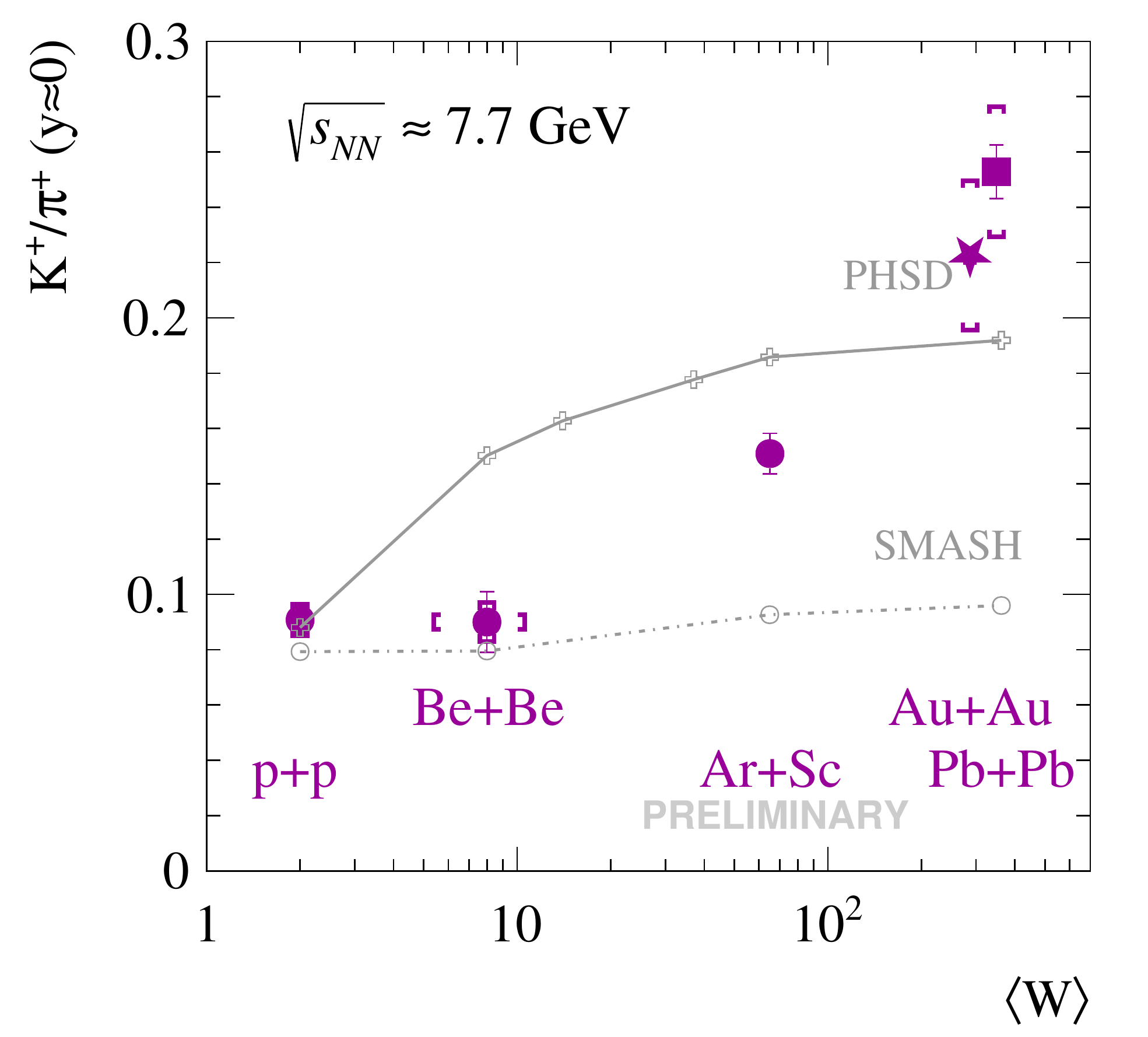}
	\includegraphics[width=0.49\textwidth]{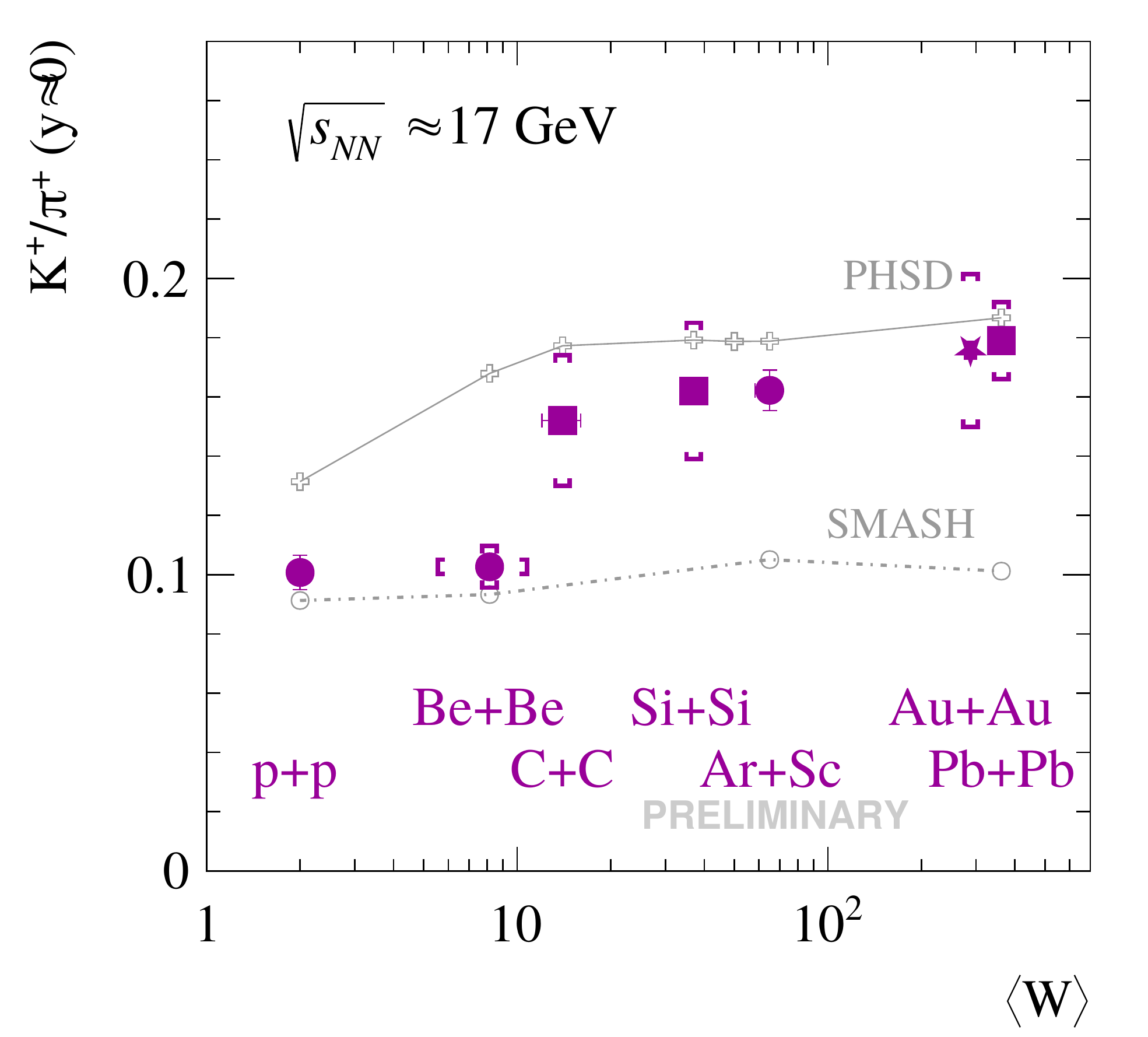}
	\caption{
	The $K^+/\pi^+$ ratio at mid-rapidity measured at  $\sqrt{s_{NN}}\approx 7.7$~GeV (\textbf{left}) and 
	$\sqrt{s_{NN}}\approx 17$~GeV (\textbf{right}) as a function of a mean number of wounded nucleons, $\langle W \rangle$, in~inelastic \textit{p+p} interactions~\cite{NA61SHINE:2017fne} and central Be+Be~\cite{NA61SHINE:2020czq}, C+C~\cite{NA49:2004jzr}, Si+Si~\cite{NA49:2004jzr}, Ar+Sc~(preliminary)~\cite{Lewicki:2020mqr,NA61SHINE:2021nye}, Au+Au~\cite{STAR:2017sal}, Pb+Pb~\cite{NA49:2007stj,NA49:2002pzu} collisions. 
	%Star points represent results form central Au+Au collisions, they were shifted by -50 units in horizontal ($\langle W \rangle$) axis for clarity of the plot. 
	Experimental results were compared with the PHSD~\cite{Cassing:2008sv,Cassing:2009vt} (open crosses) and SMASH~\mbox{\cite{Weil:2016zrk,Mohs:2019iee}} (open circles) predictions. Lines are plotted to guide the eye.
	} 
	\label{fig:system_size}
\end{figure}

Figure~\ref{fig:system_size} shows results on the $K^+/\pi^+$ ratio measured at $\sqrt{s_{NN}}\approx 7.7$~GeV (the left plot) and $\sqrt{s_{NN}}\approx 17$~GeV (the right plot) as a function of the mean number of nucleons that participated in inelastic interactions, the~so-called number of wounded nucleons $\langle W \rangle$. Since,, at high collision energies, the~ratio is weakly dependent on the collision energy, results from central Au+Au collisions at $\sqrt{s_{NN}}= 19.6$~GeV were also included in the~plot.

\unskip
 \vspace{12pt}

Let us start a discussion of model predictions concerning the system-size dependence of
the ratio from the string models.
For simplicity of the arguments, we assume that the string formation, evolution and
fragmentation is independent of $\langle W \rangle$.
The yields of $K^+$ and $\pi^+$ mesons are proportional 
to the mean number of strings, and~consequently, their ratio is independent of 
the mean number of strings, and~thus it is independent of $\langle W \rangle$.
The predictions of the SMASH model shown in Figure~\ref{fig:system_size} approximately follow this naive expectation. The~model qualitatively fails to reproduce the~data.

A different system-size dependence is predicted within statistical
models of nucleus--nucleus collisions. The~strangeness conservation imposed on the whole system leads to a fast increase of the ratio with increasing system size to its upper limit given by the grand-canonical-ensemble approximation.
The effect is referred to as canonical strangeness suppression and has been extensively studied since 1980; see, e.g.,~Refs.~\cite{Rafelski:1980gk, Redlich:1979bf,Becattini:2003ft}.
The PHSD model predictions shown in Figure~\ref{fig:system_size} show a gradual ratio increase with $\langle W \rangle$. However, in~this model, 
the change is likely to be also caused by smoothly increasing contributions from QGP 
and chiral symmetry restoration.
The PHSD model describes the main properties of the data significantly better.
However, it fails to reproduce the \emph{jump} between the results for
\emph{p+p} and Be+Be collisions and the results for heavier nuclei at $\sqrt{s_{NN}}\approx 17$~GeV; see Figure~\ref{fig:system_size}~(right).

With increasing collision energy and nuclear mass number of colliding nuclei, the~number of produced strings and their density is expected to increase. The~idea that, at~sufficiently high densities, the~strings would be close enough to interact and change their properties has been developing over the last 40 years. Many approaches have been proposed, in~particular, colour ropes~\cite{Biro:1984cf}, string fusion~\cite{Braun:1991dg,Braun:1992ss,Braun:2014ica,Braun:2015eoa,Ramirez:2020vne}, core formation~\cite{Werner:2007bf}, string melting~\cite{Lin:2004en} and
percolation~\cite{Digal:2003sg,Hohne:2005ks}.
A model that explicitly involves the rapid string-QGP changeover was proposed recently.
It is a string collapse pictured as the black hole formation using the AdS/CFT duality~\cite{Kalaydzhyan:2014zqa,Kalaydzhyan:2014tfa,Iatrakis:2015rga}.
Thus, it is natural to interpret the jump as due to a rapid changeover from strings
to QGP.
This changeover is called the 
\emph{onset of QGP fireball}.

The gradual increase of the ratio at low collision energies (see Figure~\ref{fig:system_size}~(left)) is also not reproduced by the models. 
This can be due~to the
\begin{enumerate}
\item [(i)]
Approaching equilibrium with increasing system size and evolution time;
\item [(ii)]
Weakening of the canonical strangeness suppression with increasing system size;
\item [(iii)]
Increasing role of
chiral symmetry restoration in dense hadronic matter.
\end{enumerate}

\section{Diagram of High-Energy Nuclear~Collisions}
\label{sec:diagram}

Here, for the first time, we explicitly introduce a concept of the diagram of high-energy nuclear collisions
and then, based on the experimental data and ideas discussed above, sketch its example~version.

The diagram of high-energy nuclear collisions is defined as a plot showing
domains of the dominance of different hadron-production processes in high-energy 
nuclear collisions.
The domains are indicated
in the space of laboratory-controlled parameters, the~collision energy and the nuclear-mass number of colliding nuclei.
For simplicity, we consider only central nucleus--nucleus collisions - collisions in which 
a large fraction of nucleons participated in inelastic interactions
($\langle W \rangle /$A $\approx 1$).

To sketch the example diagram, the~hadron-production processes
discussed above are~selected:
\begin{enumerate}
\item [(i)]
Creation, evolution and decay of resonances;
\item [(ii)]
Formation, evolution and fragmentation of strings;
\item [(iii)]
Creation, evolution and hadronisation of QGP.
\end{enumerate}

In addition, based on the discussion of the experimental results presented in the previous section, we assume~that
\begin{enumerate}
    \item  [(i)]
    The Pb+Pb horn locates the resonances--QGP changeover at
    $\sqrt{s_{NN}}\approx 8$~GeV;
    \item [(ii)]
    The \emph{p+p} break locates the resonances--strings changeover at
    $\sqrt{s_{NN}}\approx 8$~GeV;
    \item [(iii)]
    The jump between \emph{p+p}/Be+Be and Ar+Sc/Pb+Pb plateaus 
    locates the strings--QGP changeover at 
    $\sqrt{s_{NN}}\approx 17$~GeV;
    \item [(iv)]
    The LHC \emph{p+p} data imply QGP creation in (high multiplicity) \emph{p+p} interactions at sufficiently high (order of 1~TeV) energies.
\end{enumerate}

\begin{figure}[h]
	%\centering
	\includegraphics[width=0.80\textwidth]{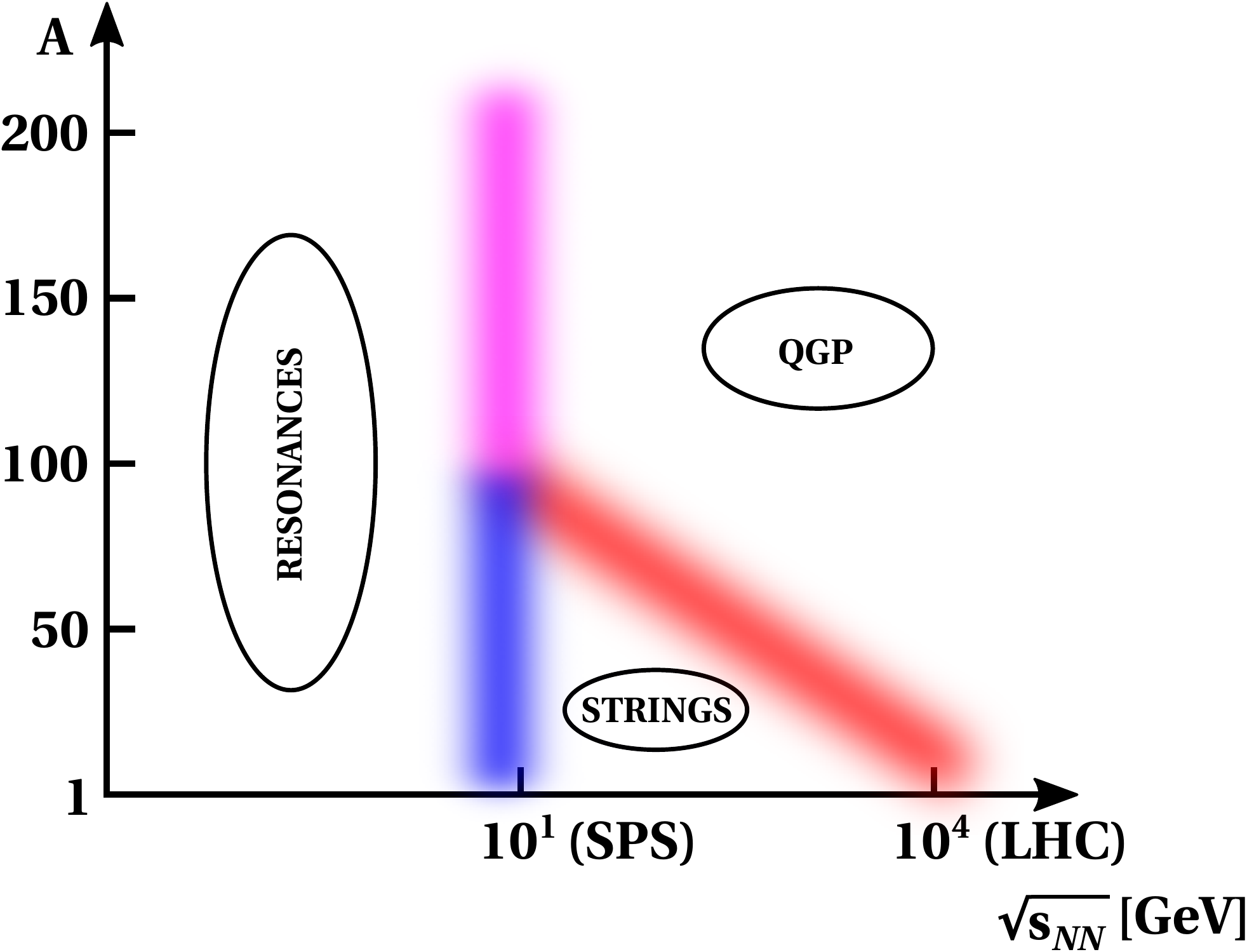}
	\caption[]
	{Schematic diagram of high-energy nuclear collisions outlined in colliding nuclei mass number, A, and~collision energy, $\sqrt{s_{NN}}$ variables. Domains in which hadron production is dominated by the creation, evolution and decay of resonances, strings
	and quark--gluon plasma are indicated as \emph{resonances}, \emph{strings} and \emph{QGP}, respectively, while thick coloured lines show the changeover regions between the domains.
	%mkuich: please keep italic
	}
	\label{fig:diagram}
\end{figure}

The diagram of high-energy nuclear collisions following these assumptions is sketched in Figure~\ref{fig:diagram}.

Two comments are in order~here.
\begin{enumerate}
\item [(i)]
The changeover resonances--strings and resonances--QGP are located at similar
collision energies ($\approx$8~GeV/c). This suggests that the resonances--QGP changeover is driven by the resonances--strings one. 
%Please confirm that the intended meaning is retained
At~high masses of colliding nuclei, 
 strings produced above at the resonances--strings changeover would have density exceeding the strings--QGP changeover. Thus the string domain disappears, and one observes direct resonances--QGP changeover. This locates the resonances--QGP changeover at the energy of the resonances--strings one.
 %MG: I modified the fragment to make it easier to understand 
\item [(ii)]
It is interesting to consider other diagrams of high-energy collisions.
Here, we discuss a simple example of
the hadron--resonance gas diagram. Hagedorn's early papers postulated that
hadrons in high-energy  collisions are produced according to
statistical thermodynamics~\cite{Hagedorn:1965st}. 
Thus, following Hagedorn's postulate, the~diagram would include only one production process - the statistical-thermodynamical production, with~Hagedorn's temperature $T_H \approx 150$~MeV. This model is clearly in contradiction with the experimental
results, as~it predicts the $K^+/\pi^+$ ratio to be independent of energy and nuclear mass number of colliding nuclei.
Over the years, the~simple Hagedorn approach evolved into many models that are much more flexible
in fitting the data; for~a recent review, see Ref.~\cite{Andronic:2017pug}.
In particular, it has been popular to fit mean hadron multiplicities, which include
multiplicities of kaons and pions, assuming that a hadron
gas in equilibrium is created at high-energy collisions.
The temperature, the~baryon chemical potential,
and the gas volume are free parameters of the model
and are fitted to the data from each reaction separately. The~model cannot predict the energy and nuclear mass dependence of hadron production in this formulation.
Thus, it is unsuitable for the diagram construction.
\end{enumerate}

%\vspace{0.2cm}
To verify the assumptions and the diagram sketched on
Figure~\ref{fig:diagram}, further analysis of the existing data and new experimental measurements
as well as the development of models is needed.
Concerning modelling,
there is a need for the development of dynamical models  
that include all three production processes.
In this paper, these models are represented 
by PHSD~\cite{Cassing:2008sv,Cassing:2009vt} which reproduces experimental
results significantly better than the SMASH model~\cite{Weil:2016zrk,Mohs:2019iee}.
The latter includes only two approaches to hadron production, resonances and strings.
Still, the~PHSD model misses important features of the experimental data shown in 
Figures~\ref{fig:horn} and~\ref{fig:system_size}.
One must reconsider the nature of the changeover between different processes to improve~predictions.

Concerning the further analysis of the existing data, 
one should extend the presented studies to other quantities which characterise hadron production in high-energy nuclear collisions. In~particular, quantities sensitive to the collective flow of matter, radial and anisotropic should be sensitive to the production mechanisms discussed. 
This important study goes beyond the scope of this introductory~paper.

Finally, concerning
the new experimental measurements, data on light and medium
%Please confirm that the intended meaning is retained
%mkuich: I confirm
mass nuclei collisions are needed---in~particular, a~precision system-size dependence to locate the strings--QGP changeover. Such a study was launched by NA61/SHINE at the CERN SPS, and~its continuation is considered in the following years~\cite{Fields:2739340,Gazdzicki:2810689}. It would be important to perform the corresponding measurements in the full range of available energies, from~the FAIR SIS-100 through NICA and SPS to CERN LHC energies. 
%They would allow for studying the processes occurring in the nuclear collisions and establishing their collision energy and system size dependence. 
In 2024, a~beam of oxygen ions is considered at the SPS and LHC in CERN~\cite{Brewer:2021kiv,Gazdzicki:2810689}, making a good start for further study. Prospects of studies with the intermediate-mass nuclear beams (e.g., Ar+Ar or Kr+Kr) at LHC energies are also vividly discussed~\cite{Citron:2018lsq}.

\begin{acknowledgments} The authors thank Mark Gorenstein and Edward Shuryak as well as members of the \NASixtyOne collaboration, in particular, Tomek Matulewicz, Andrzej Rybicki and Peter Seyboth
for discussion and comments. The authors thank  Elena Bratkovskaya, Viktar Kireyeu and Justin Mohs who helped in the calculation of predictions of the PHSD and SMASH models.
\end{acknowledgments}

\bibliographystyle{ieeetr}
\bibliography{references}

\end{document}